\documentclass{aa}
\usepackage{graphicx}
\usepackage[varg]{txfonts}
\usepackage{cases}
\usepackage{natbib}

\begin{document}
   \title{Pre-flare coronal dimmings}

   \author{Q. M. Zhang\inst{1,2}, Y. N. Su\inst{1}, and H. S. Ji\inst{1}}

   \institute{Key Laboratory for Dark Matter and Space Science, Purple
              Mountain Observatory, CAS, Nanjing 210008, China \\
              \email{zhangqm@pmo.ac.cn}
              \and
              Key Laboratory of Solar Activity, National Astronomical Observatories, CAS, Beijing 100012\\
              }

   \date{Received; accepted}
    \titlerunning{Pre-flare coronal dimmings}
    \authorrunning{Zhang et al.}

 \abstract
   {Coronal dimmings are regions of decreased extreme-ultravoilet (EUV) and/or X-ray (originally Skylab, then Yohkoh/SXT) intensities, 
   which are often associated with flares and coronal mass ejections (CMEs). The large-scale, impulsive dimmings have substantially been 
   observed and investigated. The pre-flare dimmings prior to the flare impulsive phase, however, have rarely been studied in detail.}
   {In this paper, we focus on the pre-flare coronal dimmings. We report our multiwavelength observations of the GOES X1.6 solar flare and the 
   accompanying halo CME produced by the eruption of a sigmoidal magnetic flux rope (MFR) in NOAA active region (AR) 12158 on 2014 September 10.}
   {The eruption was observed by the Atmospheric Imaging Assembly (AIA) aboard the Solar Dynamic Observatory (SDO). The photospheric line-of-sight 
   magnetograms were observed by the Helioseismic and Magnetic Imager (HMI) aboard SDO. The soft X-ray (SXR) fluxes were recorded by the GOES 
   spacecraft. The halo CME was observed by the white light coronagraphs of the Large Angle Spectroscopic Coronagraph (LASCO) aboard SOHO.}
   {About 96 minutes before the onset of flare/CME, narrow pre-flare coronal dimmings appeared at the two ends of the twisted MFR. They extended very 
   slowly with their intensities decreasing with time, while their apparent widths (8$-$9 Mm) nearly kept constant. During the impulsive and decay phases 
   of flare, typical fanlike twin dimmings appeared and expanded with much larger extent and lower intensities than the pre-flare dimmings. 
   The percentage of 171 {\AA} intensity decrease reaches 40\%. The pre-flare dimmings are most striking in 171, 193, and 211 {\AA} with 
   formation temperatures of 0.6$-$2.5 MK. The northern part of the pre-flare dimmings could also be recognized in 131 and 335 {\AA}.}
   {To our knowledge, this is the first detailed study of pre-flare coronal dimmings, which can be explained by the density depletion as a result of the 
   gradual expansion of the coronal loop system surrounding the MFR during the slow rise of the MFR.}


   \maketitle

\section{Introduction}
Flares and coronal mass ejections (CMEs) are the most violent processes of energy release in the solar atmosphere and the most important sources 
of space weather \citep{forb06,chen11,hud11,fle11}. Sometimes, the pre-flare phase, rise phase, and decay phase of flares are temporally related to the initial 
phase, impulsive acceleration phase, and propagation phase of CMEs, respectively \citep{zhang01}. Both of the phenomena are often associated with 
the eruptions of filaments \citep{zhang15,zhang16}, which are suspended by magnetic sheared arcades or magnetic flux ropes (MFRs) along the polarity inversion 
lines \citep{td99,aul10,xia14,su15}. An MFR consists of a set of magnetic field lines wrapping around its central axis and appears as a hot channel with temperature as high 
as 10 MK in the low corona \citep{zhang12a}. The large-scale eruptions usually create global disturbances propagating very long distances 
on the solar surface. For the first time, \citet{thom98} observed dimming regions and bright wavefronts with enhanced intensities that propagate quasi-radially 
from the source region of a CME. The onset of dimming is consistent with the initiation of the wavefront, i.e., coronal wave or extreme-ultraviolet (EUV) wave 
\citep{chen02,pats12}. The decrease of the intensity of dimming region is due to a decrease in plasma density rather than in temperature \citep{har00}. 
The percentage of density decrease could reach 35\%$-$40\% \citep{jin09}. The temperatures of the EUV dimmings are 1$-$4 MK \citep{zhu04,robb10,cheng12a}.
\citet{thom00} identified transient dimming regions with strong EUV emission depletion around the eruption, and the dimming areas are cospatial with the 
footprints of CMEs \citep{bew08,attr10a}. As the CME proceeds, the EUV wave propagates outwards at speed of 50$-$1500 km s$^{-1}$ and the area of trailing 
dimming increases, covering a significant fraction of the solar surface \citep{har07,thom09,attr10b,cheng12b}. Recent quadrature observations have revealed the dome-shaped 
nature of EUV waves and dimming regions \citep{pats09,vero10}. There is a special kind of dimmings that appear near the two ends of a pre-flare, \textsf{S}-shaped 
sigmoid at the beginning of the impulsive acceleration phase of CMEs \citep{mik11}. Such twin dimmings, following the eruption of a twisted 
MFR, can also persist for tens of hours to more than three days \citep{ster97,kah01}. The decrease of twin dimming intensity is correlated with the increase of flaring 
arcade intensity in EUV wavelengths \citep{zar99}. Interestingly, twin dimming regions are found to rotate around the center of the flare site due to rotation 
of the erupting filament \citep{mik11}. Similar to typical coronal dimmings that have a circular or elliptical shape, twin dimmings also originate from density depletion 
as a result of magnetic field line stretching or reconfiguration associated with CMEs. Spectroscopic observations from space telescopes have revealed strong 
plasma upflows at speeds of tens to hundreds of km s$^{-1}$ in the coronal dimming regions \citep[e.g.,][]{har01,doll11}. The gradual recovery of dimming regions 
lasts for a long time by re-establishment of the bright coronal loops \citep{rein08,attr10a}. 
Sophisticated numerical simulations have also improved our understandings of coronal waves and dimmings. 
\citet{ima07} discovered that the velocities (16$-$160 km s$^{-1}$) of upflow in the dimming region depend on the formation temperatures of the emission lines, i.e., 
hotter lines show faster upflow velocities. \citet{ima11} found that the temperature-dependent upflow in the dimming region can be well modeled by quasi-steady 
flow in a vertical flux tube whose cross section expands super-radially with height from the solar surface.

Since the launch of Solar Dynamic Observatory (SDO), coronal dimmings following CMEs have been extensively investigated \citep[][and references therein]{liu14,war15}.
However, the pre-flare or pre-eruption dimmings have rarely been reported. \citet{cheng16} studied the coronal dimmings on 2011 December 26 in detail 
and found that the rapid dimmings started after the onset of fast magnetic reconnection and CME acceleration. At the end of the paper, they mentioned that, 
at some locations, the gradual dimmming started about 30 minutes before the CME eruption and impulsive flare reconnection. The authors proposed that 
the gradual and weak pre-eruption dimming may reflect the slow expansion of coronal structures. 
In this paper, we report long-term EUV dimmings prior to the onset of flare and CME observed 
by the Atmospheric Imaging Assembly \citep[AIA;][]{lem12} aboard SDO. On 2014 September 10, an inverse-{\sf S}, sigmoidal MFR formed in NOAA 
active region (AR) 12158 close to the disk center. At $\sim$17:21 UT, the MFR erupted and produced a GOES X1.6 flare that peaked at $\sim$17:45 UT 
and a full halo CME that propagated towards the earth \citep{li15a,li15b,li15c,cheng15,zhao16,dud16}.
The rest of this paper is structured as follows. Data analysis is described in Sect.~\ref{s-data}, and the results are shown in Sect.~\ref{s-res}.
Discussion about the nature and significance of pre-flare dimmings are presented in Sect.~\ref{s-disc}. Finally, we give a summary of the results in Sect.~\ref{s-sum}.

\section{Data analysis} \label{s-data}
The eruption was clearly observed by AIA, which has seven EUV filters (94, 131, 171, 193, 211, 304, and 335 {\AA}) with a cadence of 12 s and two UV filters (1600 
and 1700 {\AA}) with a cadence of 24 s. Line of sight (LOS) magnetograms from the Helioseismic and Magnetic Imager \citep[HMI;][]{sch12} aboard SDO with a cadence of 45 s 
were used for studying the photospheric magnetic field of the AR. The full-disk EUV images and magnetograms have spatial resolutions of 1$\farcs$2 and 1$\arcsec$, respectively.
The AIA and HMI level\_1 fits data were calibrated using the standard \textit{Solar Software} (\textit{SSW}) programs \textit{aia\_prep.pro} and \textit{hmi\_prep.pro}. 
The fast halo CME\footnote{http://cdaw.gsfc.nasa.gov/CME\_list/UNIVERSAL/2014\_09/} at a linear speed of 1267 km s$^{-1}$
was observed by the C2 and C3 white light (WL) coronagraphs of the Large Angle Spectroscopic Coronagraph \citep[LASCO;][]{bru95} aboard SOHO. 
The LASCO/C2 images were calibrated using the \textit{SSW} program \textit{c2\_calibrate.pro}. The high-cadence (2.047 s) SXR fluxes of the flare were recorded by GOES.

\section{Results} \label{s-res}
\begin{figure}
\centering
\includegraphics[width=8cm,clip=]{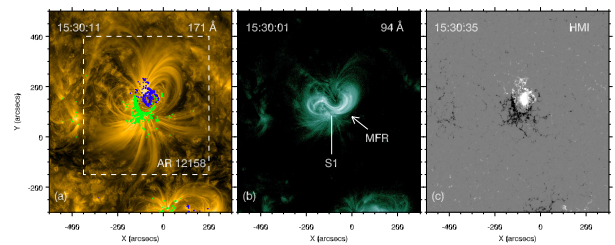}
\caption{(a) AIA 171 {\AA} image at 15:30:11 UT. The AR 12158 is enclosed by the white dashed box. 
(b) AIA 94 {\AA} image at 15:30:01 UT, which features an MFR as pointed by the white arrow. 
A vertical slice S1 (white solid line) is used for investigating the slow rise and expansion of the coronal loops. 
(c) HMI LOS magnetogram at 15:30:35 UT. The contours ($\pm$400, $\pm$800, and $\pm$1200 G) of the LOS 
positive and negative magnetic field are overlaid on the 171 {\AA} image with blue and green lines, respectively.}
\label{fig1}
\end{figure}

In Fig.~\ref{fig1}, panel (a) shows the 171 {\AA} image at 15:30:11 UT prior to the onset of flare. The AR 12158 is enclosed by the white dashed box 
(550$\arcsec\times$550$\arcsec$). Panel (b) shows the 94 {\AA} image at 15:30:01 UT with much higher formation temperature ($\sim$6 MK). 
The image features the bright, inverse-{\sf S}, sigmoidal MFR embedded in the core of AR \citep{cheng15}. 
The photospheric LOS magnetogram observed by HMI at 15:30:35 UT is displayed in panel (c). The AR was associated with a complex $\beta\gamma\delta$ sunspot. 
The contours ($\pm$400, $\pm$800, and $\pm$1200 G) of the photospheric LOS magnetic field of the positive and negative polarities are 
overlaid on the 171 {\AA} image with blue and green lines, respectively. It is clear that the MFR resides between the positive and negative polarities.

\begin{figure}
\centering
\includegraphics[width=8cm,clip=]{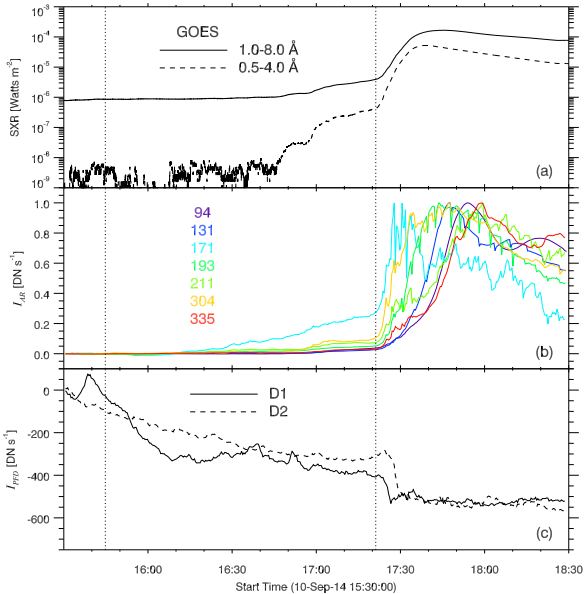}
\caption{(a) GOES SXR fluxes of the flare during 15:30$-$18:30 UT. (b) Temporal evolutions of the normalized, integral intensities of the AR within the dashed box of Fig.~\ref{fig1}(a). 
(c) Temporal evolutions of the 171 {\AA} base-difference intensities of D1 and D2 marked in Fig.~\ref{fig5}(b). \textit{PFD} stands for pre-flare dimming.
The two vertical dotted lines in each panel denote the start time (15:45 UT) and end time (17:21 UT) of pre-flare dimmings.}
\label{fig2}
\end{figure}

In Fig.~\ref{fig2}, the GOES SXR light curves in 0.5$-$4.0 {\AA} and 1$-$8 {\AA} are plotted in panel (a) with dashed and solid lines. The SXR 
fluxes increase gradually from $\sim$15:45 UT and rise rapidly during the impulsive phase of the flare before decreasing slowly during the long decay phase \citep{li15a}. 
We also calculated the integrated, base-difference EUV intensities of the whole AR within the white box of Fig.~\ref{fig1}(a). The temporal evolutions of the 
normalized intensities during 15:30$-$18:30 UT are depicted with colored lines in Fig.~\ref{fig2}(b). It is clear that the total EUV intensities of the AR increase very 
slowly during the pre-flare phase (15:45$-$17:21 UT), which is similar to the SXR light curves. The slow rises of both SXR and EUV fluxes of the AR suggest 
that plasma heating as a result of the release of magnetic free energy is already taking place, presumably via magnetic reconnection \citep{cheng15}. 
The early heating is also supported by the fact that the 94 and 131 {\AA} light curves with higher peak formation temperatures do not start increasing until later.

\begin{figure}
\centering
\includegraphics[width=8cm,clip=]{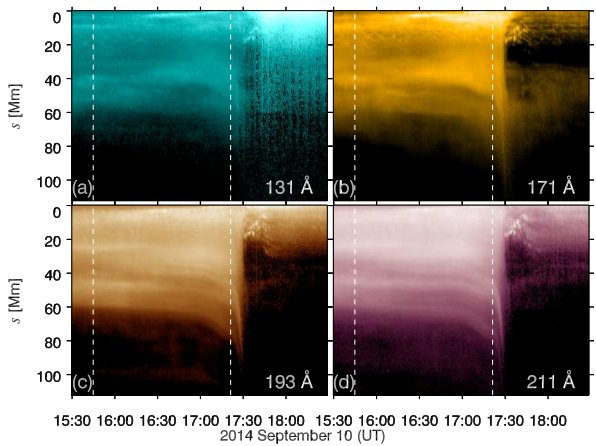}
\caption{Time-slice diagrams of S1 in four AIA filters using the original images. $s=0$ Mm represents the flare site. 
The two vertical dashed lines in each panel signify the start time (15:45 UT) and end time (17:21 UT) of pre-flare dimmings.}
\label{fig3}
\end{figure}

In Fig.~\ref{fig1}(b), we draw a vertical slice (S1), which originates from the flare site with a length of 112 Mm.
The temporal evolutions of the intensities along S1 in four of the AIA filters during 15:30$-$18:30 UT are displayed as time-slice 
diagrams in Fig.~\ref{fig3}. It is clear that the coronal loops are undergoing slow rise and expansion during the mild heating of the AR before the onset of flare, 
especially in 193 and 211 {\AA}.

\begin{figure}
\centering
\includegraphics[width=8cm,clip=]{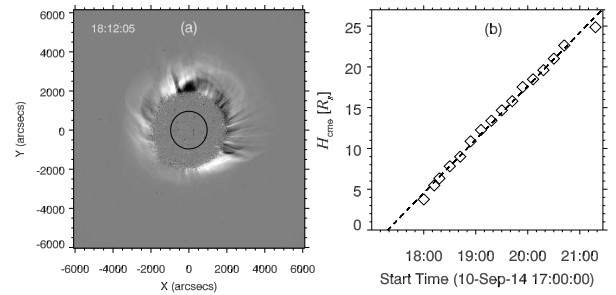}
\caption{(a) WL base-difference image observed by LASCO/C2 at 18:12:05 UT. (b) Height-time profile of the halo CME. 
The dashed line represents the least-square linear fitting of the profile.}
\label{fig4}
\end{figure}

The halo CME appeared and expanded isotropically in the field-of-view of LASCO/C2 during 18:00$-$18:36 UT. In Fig.~\ref{fig4}, the base-difference WL image 
of the CME at $\sim$18:12 UT is displayed in the left panel. The height-time profile of the CME is plotted in the right panel with diamonds. The result of least-square fitting 
of the profile is overlaid with a dashed line. The extrapolated initiation time of CME coincided with the onset of flare at $\sim$17:21 UT.

\begin{figure}
\centering
\includegraphics[width=8cm,clip=]{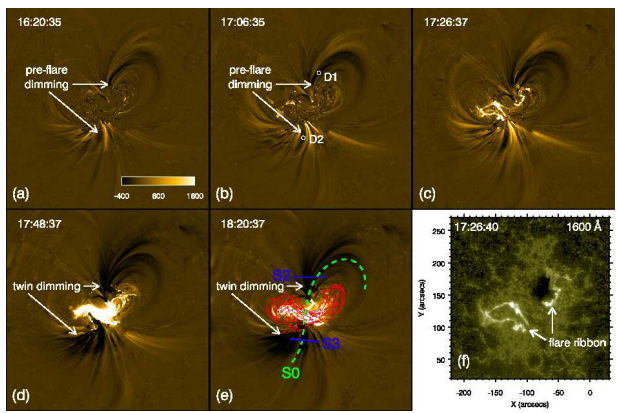}
\caption{(a)-(e) Snapshots of the base-difference images in 171 {\AA}. The pre-flare dimmings and twin dimmings are pointed by the white arrows.
In panel (b), two circles mark the locations of two representative pre-flare dimmings (D1 and D2).
In panel (e), a curved slice S0 (green dashed line) is used for investigating the temporal evolution of the coronal dimmings. 
The intensity contours of the 94 {\AA} image in Fig.~\ref{fig1}(b) are overlaid with red lines. 
(f) AIA 1600 {\AA} image at 17:26:40 UT that features two bright flare ribbons. 
The temporal evolution of the flare is shown in a movie (\textit{flare.avi}) available in the online edition.}
\label{fig5}
\end{figure}

Taking the EUV images at $\sim$15:30 UT as base images, we obtained the subsequent base-difference images until 18:30 UT (see the online movie \textit{flare.avi}). 
In Fig.~\ref{fig5}, panels (a)-(e) demonstrate five snapshots of the base-difference images in 171 {\AA}. It is clear that narrow dimming regions appeared at the 
northern and southern ends of the MFR before flare, which are denoted by ``pre-flare dimming" in panels (a) and (b). As time goes on, the pre-flare dimmings 
extended, especially for the northern part in panel (b). During the impulsive phase of flare when the MFR erupted out of the corona, two bright ribbons showed up 
in the 171 {\AA} and 1600 {\AA} images (see panels (c) and (f)). Meanwhile, the areas of dimming regions expand rapidly with their intensities decreasing to a great 
extent, forming the typical fanlike ``twin dimming" indicated in panels (d) and (e). The lack of bright loops nearby in the base-difference images suggests that 
this is most likely a density depletion. During the decay phase of flare, the dimmings sustained and extended very slowly.
In panel (e), the 94 {\AA} intensity contours are superposed on the 171 {\AA} image with red lines. It is evident that both the pre-flare dimmings and twin dimmings 
originated at the two ends of the sigmoid. This is consistent with the finding that the observed coronal dimmings are located around the footpoints of a erupting flux rope \citep{su11}.

\begin{figure}
\centering
\includegraphics[width=8cm,clip=]{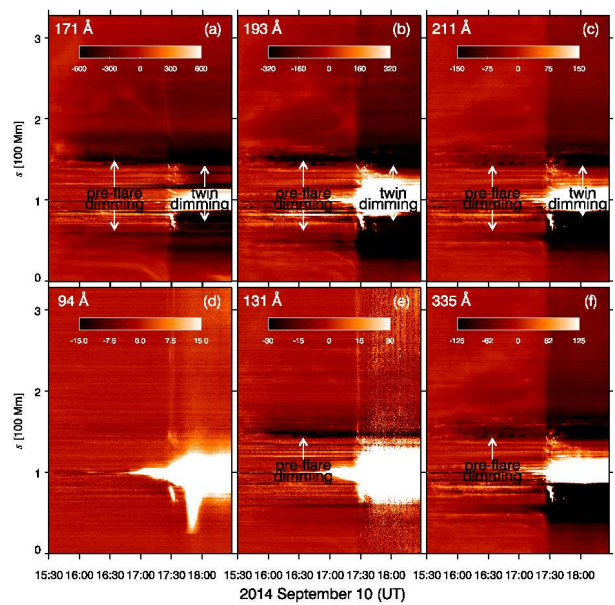}
\caption{Time-slice diagrams of S0 in 171, 193, 211, 94, 131, and 335 {\AA} using the AIA base-difference images. 
$s=0$ and $s=327.5$ Mm represent the southeastern and northwestern endpoints of S0. 
The dark pre-flare dimmings and twin dimmings are pointed by the white arrows in each panel.}
\label{fig6}
\end{figure}

To illustrate the temporal evolution of the dimmings more clearly, we selected a curved slice (S0 with dashed green line) that passes through the core of AR and 
dimming regions in Fig.~\ref{fig5}(e). The total length of S0 is 327.5 Mm. The evolutions of the base-difference intensities in 171, 193, and 211 {\AA} 
along S0 are displayed as time-slice diagrams in the top panels of Fig.~\ref{fig6}. Both the pre-flare dimmings and twin dimmings are distinct 
as dark regions in the diagrams, especially in 171 {\AA}, indicating that the plasma dominantly involved in the dimmings was in the range of 0.6$-$2.5 MK. 
The pre-flare dimmings started at $\sim$15:45 UT and lasted for $\sim$96 minutes until the beginning of the flare impulsive phase at $\sim$17:21 UT. 
Afterwards, twin dimmings showed up and expanded rapidly with much larger extent and lower intensities. The percentage of intensity decrease is $\sim$40\%.
It is noticed that the northern part of the pre-flare dimmings was always more striking and evident than the southern 
part, while the southern part of the twin dimmings was always stronger than the northern part during the flare. The evolutions of the base-difference intensities in 94, 131, and 
335 {\AA} along S0 are displayed as time-slice diagrams in the bottom panels of Fig.~\ref{fig6}. Only the northern part of the pre-flare dimmings could be identified 
with slightly lower magnitudes than those in the cooler filters, while the southern part is hardly visible in these wavelengths.

To compare the pre-flare dimming regions with the whole AR, we selected two representative positions (D1 and D2) in the pre-flare dimming regions, which are labeled 
with small circles in Fig.~\ref{fig5}(b). The temporal evolutions of the base-difference intensities of D1 and D2 in 171 {\AA} are displayed in Fig.~\ref{fig2}(c) with 
solid and dashed lines, respectively. It is obvious that the intensities of D1 and D2 decrease gradually from $\sim$15:45 UT until 17:27 and 17:30 UT, respectively. 

\section{Discussion} \label{s-disc}
\subsection{The nature of pre-flare dimmings} \label{s-nature}
Despite of extensive observations and investigations in the past, the origins of coronal dimmings are not completely understood. \citet{mas14} summarized the mechanisms of 
coronal dimmings, including the mass-loss dimmings as a result of ejection of emitting plasma with temperatures of several MK \citep{har00}, thermal dimmings 
because different EUV filters have different formation temperatures \citep{cheng11} and Doppler dimming, to name a few. In our study, the large-scale twin dimmings 
followed the pre-flare dimmings. It has been widely accepted that the twin dimmings originate from the density depletion as a result of eruption of an MFR \citep{su11}. 
Before the flare impulsive phase, there was no clear signature of mass loss. The pre-flare dimmings result from density depletion when the whole coronal loop system
surrounding the MFR undergoes gradual volume expansion during the slow rise of the MFR. Such a precursor expansion in this event has also been noticed by \citet{dud16}. 
As illustrated in the schematic cartoon of \citet{cheng15}, the tether-cutting magnetic reconnection occurred during the pre-flare phase \citep{moo01}, which 
has two effects. One is the mild heating of the MFR residing in the core of AR (see Fig.~\ref{fig2}(a)). The other is the decrease of constrain on the MFR from the envelop 
magnetic field as well as the increase of poloidal magnetic flux of the MFR at the same time, leading to the slow rise of the MFR \citep{ster05,chi07,ster11}. 
This is usually accompanied by gradual volume expansion since the magnetic field and electron number density decrease with height in the low-$\beta$ corona.

\begin{figure}
\centering
\includegraphics[width=8cm,clip=]{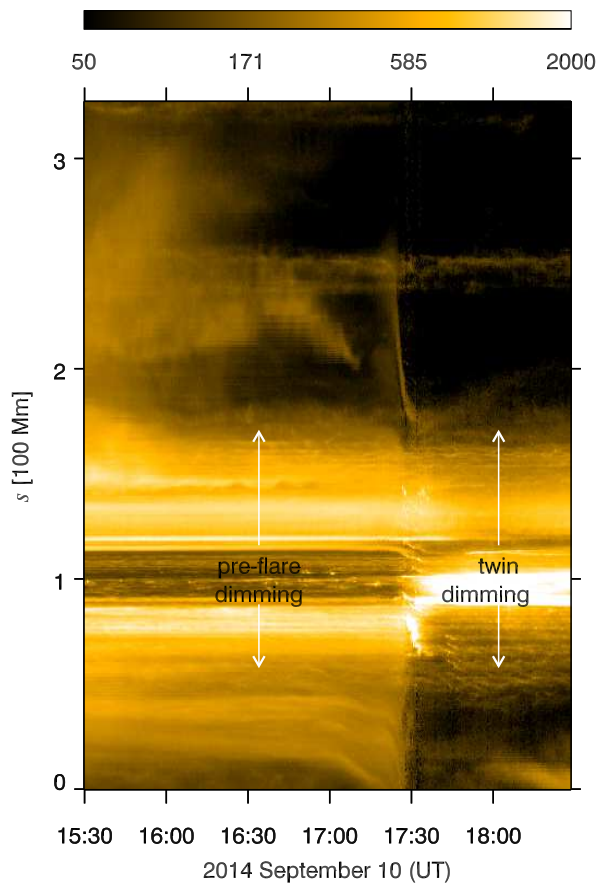}
\caption{Time-slice diagram of S0 derived from the original 171 {\AA} images instead of the base-difference images.
$s=0$ Mm represents the southeastern endpoint of S0. The dark pre-flare dimmings and twin dimmings are pointed by the arrows.}
\label{fig7}
\end{figure}

Considering that both pre-flare dimmings and twin dimmings are clearly demonstrated in the base-difference images, there is a concern that the pre-flare dimmings are 
artifact of the base-difference image processing. To dispel this doubt, we make time-slice diagram of S0 using the original images in 171 {\AA} instead of the base-difference 
images. It is obvious from Fig.~\ref{fig7} that both pre-flare dimmings and twin dimmings are distinguishable as dark regions. The locations and time of the dimmings are the same 
as those in Fig.~\ref{fig6}(a). Therefore, the pre-dimmings are real and convincing, not artifact of the base-difference image processing.

\begin{figure}
\centering
\includegraphics[width=8cm,clip=]{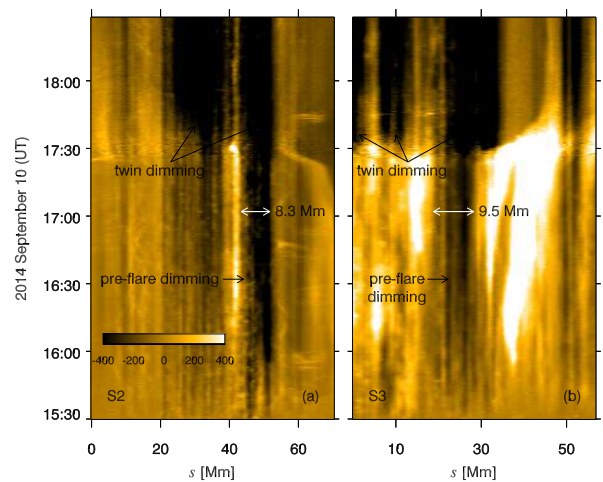}
\caption{Time-slice diagrams of S2 (left panel) and S3 (right panel) in 171 {\AA} using the AIA base-difference images. 
$s=0$ Mm represents the left endpoints of S2 and S3. The dark pre-flare dimmings and twin dimmings are pointed 
by the black arrows in each panel. The white double-headed arrows denote the apparent widths of the pre-flare dimmings along S2 and S3.}
\label{fig8}
\end{figure}

Another concern is that the pre-flare dimmings might result from the 
transverse shifts of the bright, adjacent coronal loops before its impulsive eruption of the MFR. To test this conjecture, we selected two short, straight slices (S2 and 
S3 with solid blue lines) that cross the northern and southern parts of dimming regions in Fig.~\ref{fig5}(e). The total lengths of S2 and S3 are 70.6 and 57.3 Mm.
The evolutions of the base-difference intensities in 171 {\AA} along S2 and S3 are displayed as time-slice diagrams in the left and right panels of Fig.~\ref{fig8}.
Note that the time axis and distance axis are different from those of Fig.~\ref{fig6}. Starting from $\sim$15:45 UT, the northern part of pre-flare dimmings 
decreases in intensity, with its apparent width ($\sim$8.3 Mm) nearly keeping constant. After the onset of flare, the northern part of twin dimmings expanded rapidly 
to a much broader extent with lower intensity (see panel (a)). Similar to the northern part, the southern part of pre-flare dimmings decreases in intensity with the 
apparent width ($\sim$9.5 Mm) keeping constant before the appearance of the southern part of twin dimmings (see panel (b)). There is no clear signature of coronal 
loop shifts in the transverse directions before the impulsive phase of flare.

\subsection{The significance of pre-flare dimmings}
Although it is not easy to predict when a filament (flux rope) erupts, which potentially leads to a flare and/or a CME, there are precursor activities a few to tens of minutes before the eruptions, 
such as initial heating and brightenings as a result of magnetic reconnection \citep{ster05,jos11}, strong blue asymmetry in the H$\alpha$ line \citep{cho16}, radio noise storm 
due to reconstruction of the large-scale magnetic structure \citep{lan81}, filament oscillation \citep{zhang12b}, to name a few. In this paper, we observed long-term ($\sim$96 minutes) 
coronal dimmings before the onset of the impulsive phase of the flare as well as the initiation of the halo CME, which can be considered as another precursor of solar eruptions. 
Additional case studies and statistical works using multiwavelength observations are worthwhile to investigate the nature of pre-flare coronal dimmings. 
If the proposal is consolidated, it will be undoubtedly valuable for the space weather prediction.

\section{Summary} \label{s-sum}
In this paper, we report our multiwavelength observations of the flare/CME event that occurred in AR 12158 as a result of eruption of an MFR on 2014 September 10, 
focusing on the study of pre-flare coronal dimmings. The main results are summarized as follows:

\begin{enumerate}
\item{The pre-flare dimmings appeared at the two ends of the twisted MFR and lasted for $\sim$96 minutes until the onset of flare and initiation of the halo CME. 
The narrow pre-flare dimmings extended very slowly with their intensities decreasing with time. Their apparent widths (8$-$9 Mm), however, did not change a lot.}
\item{The pre-flare dimmings are observed by SDO/AIA mainly in 171, 193, and 211 {\AA} with formation temperatures of 0.6$-$2.5 MK. 
The northern part of pre-flare dimmings could be identified in 131 and 335 {\AA} with higher formation temperatures.}
\item{During the impulsive and decay phases of flare, typical fanlike twin dimmings appeared and expanded with much larger extent and lower intensities than the pre-flare dimmings. 
The percentage of 171 {\AA} intensity decrease reaches $\sim$40\%. It is likely that the pre-flare dimmings are progenitor of the large-scale twin dimmings.}
\item{We conclude that the pre-flare dimmings originate from density depletion as a result of the gradual expansion of the coronal loop system surrounding the MFR 
during the slow rise of the MFR. Our findings provide strong supports for the conjecture of \citet{cheng16} and have potential significance to space weather prediction.}
\end{enumerate}

\begin{acknowledgements} 
The authors sincerely appreciate the referee for detailed and valuable comments. We also thank J. Zhang, M. D. Ding, P. F. Chen, R. Liu, B. Kliem, G. Aulanier, 
and X. Cheng for fruitful discussions. SDO is a mission of NASA\rq{}s Living With a Star Program. AIA and HMI data are courtesy of the NASA/SDO science teams. 
SOHO is a project of international cooperation between ESA and NASA. 
QMZ is supported by the Surface Project of Jiangsu No. BK20161618, NSFC No. 11303101, 11333009, 11573072, and the 
open research program of Key Laboratory of Solar Activity, National Astronomical Observatories, CAS No. KLSA201510. 
HSJ is supported by the Strategic Priority Research Program--The Emergence of Cosmological Structures of the CAS, Grant No. XDB09000000. 
YNS is supported by NSFC 11473071, Youth Fund of Jiangsu BK20141043, and the One Hundred Talent Program of Chinese Academy of Sciences.
\end{acknowledgements}

\end{document}